\documentclass[9pt,twocolumn,twoside]{osajnl}
\usepackage{caption}
\usepackage{subcaption}
\usepackage{multirow}

\journal{optica} 

\setboolean{shortarticle}{true}

\title{Efficient ($\sim$10$\%$) Generation of Vacuum Ultraviolet Femtosecond Pulses via Four-Wave Mixing in Hollow-Core Fibers}

\author[1]{Ruaridh Forbes}
\author[2]{Paul Hockett}
\author[3]{Quentin Leterrier}
\author[2,*]{Rune Lausten}

\affil[1]{Linac Coherent Light Source, SLAC National Accelerator Laboratory, Menlo Park, CA 94025, USA}
\affil[2]{National Research Council of Canada, 100 Sussex Drive, Ottawa, ON K1A 0R6, Canada}
\affil[3]{École Normale Supérieure de Lyon, 15 parvis René Descartes, 69342 Lyon, France}

\affil[*]{Corresponding author: rune.lausten@nrc-cnrc.gc.ca}

\begin{abstract}
We report the generation of the 5th harmonic of Ti:sapphire, at 160~nm, with more than 4~$\mu$J of pulse energy, and a pulse length of 37~fs with a 1~kHz repetition rate. The vacuum ultraviolet pulses are produced using four-wave difference frequency mixing in a He-filled stretched hollow-core fiber, driven by a pump at 267~nm and seeded at 800~nm. Guided by simulations using Luna.jl, we are able to optimize the process carefully. The result is a conversion efficiency of $\sim$10$\%$ from the 267~nm pump beam, rivaling efficient optical mixing schemes in nonlinear crystals. 
\end{abstract}

\setboolean{displaycopyright}{true}

\begin{document}

\maketitle

The generation of femtosecond vacuum ultraviolet (VUV, 100-200~nm) pulses remains difficult~\cite{Chapman2014f}, but has potential for a wide range of applications, including in ultrafast optics, molecular spectroscopy and dynamics~\cite{Eikema2011, Chergui2014b, Reid2016b, schuurman2022TimeresolvedPhotoelectronSpectroscopy}. 
However, VUV conversion efficiencies have so far been quite low, in the $\sim0.1\%$ range even in the most efficient cases based on four-wave mixing (FWM)~\cite{Noack_2muJ, forbes2022EfficientGeneration7th}, naturally limiting their adoption. 
In the deep UV (DUV) range, efficient conversion has been demonstrated through FWM in argon filled hollow-core fiber (HCFs)~\cite{Jailaubekov_2005}, with $\sim$30$\%$ at 267~nm and $\sim$1-10$\%$ between 220-240~nm.  In argon filled kagomé-style
photonic crystal fiber, an impressive 38$\%$ was converted to 270~nm through FWM~\cite{belli_2019}.
Recent progress in the understanding of optical soliton dynamics has provided another approach through tunable resonant dispersive wave emission, where $\geq 5\%$ conversion to 170~nm, and $\geq 3\%$ conversion to 160~nm,  with bandwidths corresponding to few-femtosecond pulses in the VUV, has been demonstrated~\cite{Travers2019}.
In the following we build on foundational work on nonlinear optics in HCFs~\cite{Durfee_1997,Durfee_1999,Durfee_2002,Misoguti_2001,Tzankov_2007}, taking advantage of the high pressure phase matching of the third order four-wave difference frequency mixing (FWDFM) process in He-filled stretched HCF~\cite{Nagy_2008}.
The FWDFM process process involves two pump-pulses at $\omega_p$ and a seed-pulse at $\omega_s$ to generate a VUV idler-pulse at $\omega_i=2\omega_p-\omega_s$. 
The core of this technique is two physical characteristics of HCFs. Firstly, they provide tight field-confinement over an extended interaction length, making it possible to take advantage of the smooth dispersion profiles of noble gasses and, in spite of their relatively low non-linearity, achieve efficient frequency conversion.
Secondly, HCFs also enable phase-matching for the DFFWM process at a specific pressure, where the positive dispersion of the gas is balanced by the negative contribution from wave-guide propagation.
%
%
The result is very efficient frequency mixing. Pumping at 267~nm and seeding at 800~nm, we demonstrate $\sim$10$\%$ conversion of pump into the VUV, rivalling efficient nonlinear optical mixing schemes in nonlinear optical crystals. This evolution opens the door to scaling these types of VUV sources to higher repetition rate laser systems with lower peak power.

\begin{figure}[t]
  \centering
     \includegraphics[width=0.4\textwidth]{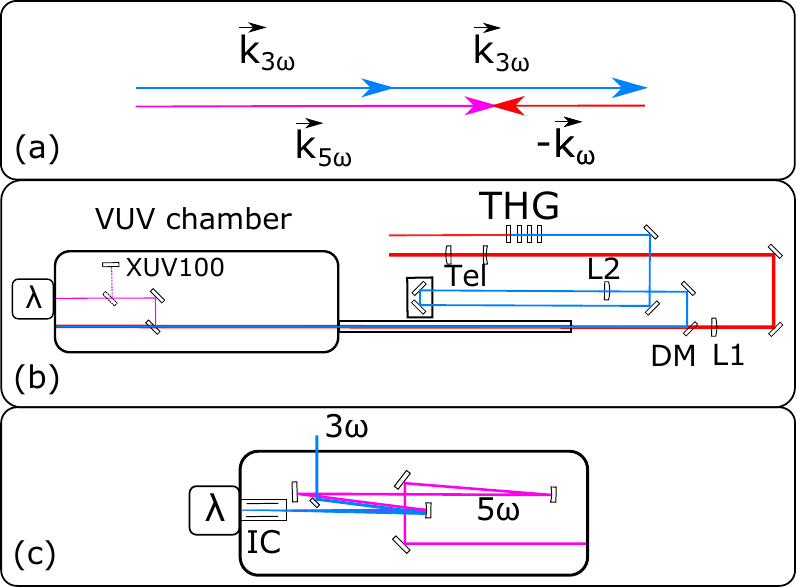}
  \caption{\label{fig:5w_generation}(a) Wave vector diagram for the phase matching between the $\omega$ and 3$\omega$ beams in the FWDFM process. (b) Experimental setup for the HCF fiber generation of the fifth harmonic. The fundamental, and third harmonic beams are recombined on a dichroic mirror (DM), mode-matched to the fiber EH$_{11}$ mode. L1 is a thin fused silica lens with f=500~mm, and L2 is a thin CaF$_{2}$ lens with f=1358~mm. The VUV chamber and fiber chambers are both filled with He to the phase matching pressure of 45.5~kPa. The fifth harmonic is separated from the driving beams with dielectric mirrors, and either sent to the VUV spectrometer (VS7550), or to a VUV photodiode (XUV100), with a flip mirror. (c) The setup for measuring the pulse length of the VUV pulse, though two-photon ionization cross-correlation in a Xe-filled ionization cell (IC). Details are in the main text.}
\end{figure}

In the work presented herein, there were two key parts to achieving a major improvement in conversion efficiency. The first an experimental milestone of using stretched HCFs, following the guidelines presented in the work of Nagy {\it et al.}~\cite{Nagy_2008}. This enables the construction of fused silica HCFs with near perfect properties, and transmission very close to theory predictions in a simple and reproducible manner~\cite{Marcatili}. 
The second key was the use of the modeling software, Luna.jl~\cite{Luna}, which enabled the exploration of the large parameter space for optimizing FWM signal generation in HCFs. A schematic of the experimental setup can be seen in Fig.~\ref{fig:5w_generation}.
The pump laser is a standard Chirped Pulse Amplification Ti:sapphire laser, with a pulse length of 35~fs (1~kHz Coherent Legend Elite DUO USP).
Starting with 2.7~mJ from this system, we first split off $\sim$0.2~mJ to act as the seed in the FWM process with a 90/10 beamsplitter. We then split the remaining $\sim$2.4~mJ on a 65/35 beamsplitter, to get 1.6~mJ for third harmonic generation (THG, 3$\omega$). The tripler consists of second harmonic generation (SHG) stage, using BBO type-1, 10x10x0.2mm, $\theta$=29.2deg, converting ~30$\%$ of P-polarized $\omega$ to S-Polarized 2$\omega$. This then propagates through a calcite crystal, which compensates for the group delay between the $\omega$ and 2$\omega$. After this the $\omega$ and 2$\omega$ pulses are rotated $\lambda$ and $\lambda$/2 respectively, to bring them both into S, using a zero-order dual-wavelength waveplate. Finally 3$\omega$ is generated P-polarized in another BBO crystal, via type-1 phase-matching, 10x10x0.1mm, $\theta$=44.3deg. 
The power of the seed can be varied by a half waveplate, thin film polarizer (TFP) setup, and the power of the 3$\omega$ pump can be varied by adjusting the half-wave plate in the frequency tripling scheme. The power of the seed and the pump were typically varied between 0 and 45~$\mu$J. The upper limit set in order to avoid the onset of nonlinear effects in the input window. The seed and the pump are coupled to a 0.5~m long stretched HCF~\cite{Nagy_2008}, with an ID=150~$\mu$m (Polymicro, TSP150665), through a 0.5~mm MgF$_2$ entrance window (Crystran, VUV grade). The fiber can be evacuated and filled with the noble gas He. Experimentally achieved coupling efficiencies were 60$\%$ and 89$\%$ for the pump and seed, respectively.
%
After the fiber, the VUV pulse is separated from the pump and seed with dielectric mirrors (Layertec, 45~deg, 160~nm HR), and either sent to a spectrometer (Resonance, VS7550 VUV/UV mini spectrometer), or to a VUV photodiode (OSI Optoelectronics, XUV100), which allowed us to measure the VUV power without significant background after suitable baffling. Calibration of the VUV photodiode was done by measuring the 3$\omega$ with a calibrated thermal head (3A-FS, Ophir), and then using that to calibrate the response of the photodiode at this wavelength. This was subsequently corrected for the difference in the quantum efficiency at 5$\omega$ versus 3$\omega$. In the quoted results from the power measurements, we correct for the reflectivity of the dielectric, and VUV aluminum mirrors (Acton Optics, ($\#$1200, Al + MgF$_2$)), in the beam path, using reflectivity curves from the manufacturer. The generated VUV pulses from the capillary at optimal phase matching pressure showed good stability. Short term we see VUV pulse energy fluctuations of $\leq$5$\%$, and the average VUV pulse energy for longer intervals remains constant, as long as the pointing stability into the capillary is stable, and the setup does not have any leaks, allowing oxygen into the VUV beam path. 
Considering that the wave vectors mismatch, $\Delta k$, of the FWDFM process, can be written as $\Delta k = 2k_{3\omega}-k_\omega-k_{5\omega}$, and inserting the expressions for the contributions to the wave vectors from the dispersion of the noble gas, at pressure $P$, and the modal dispersion from the capillary propagation, based on the Marcatili and Schmeltzer paper \cite{Marcatili}, the wave vector mismatch splits into a positive, pressure dependent, gas term, and a negative modal propagation term $\Delta k = P \Delta k_{gas}-\Delta k_{mode}$ \cite{Durfee_2002}. 
To characterize the pressure dependence of the 5$\omega$ conversion efficiency, pulse energy measurements of the VUV were performed as a function of He pressure, for set pump and seed powers. The power scaling of the VUV generation was also investigated by varying the seed and pump power, while maintaining the fiber at the optimal phase matching pressure.
The experimental measurements may be compared with the simple theory expression: 
\begin{equation}
I_{5\omega} \propto  N^2 \mid {\chi}^{(3)} \mid^{2} \frac{I_{\omega} I_{3\omega}^2 L^2 ~\mbox{sinc}^2(\Delta k L/2)}{n_{\omega} n_{3\omega}^2 \lambda_{5\omega}^2 } \\
\label{eq:simple-PM}
\end{equation}
Previous work from Noack \textit{et al.} utilized this expression to obtain excellent agreement with FWM experiments to generate the fifth harmonic (5$\omega$) of Ti-Sapphire \cite{Noack_2muJ}.
It contains the phase-mismatch, ${\Delta}k$, $N$ (which is proportional to the pressure, $P$) is the number density  of the He atoms, ${\chi}^3$ is the third-order nonlinear coefficient, $I$ is the intensity, $L$ is the interaction length, and $n$ is the refractive index.  
The experimental power-scaling measurements agree very well with the expectations based on Eq.~\ref{eq:simple-PM}, demonstrating linear scaling with the seed and quadratic with the pump, over the range explored, which was limited by nonlinear effects, such as self-focusing and color center formation in the input window. In the current experiments, pump and seed powers up to 45~$\mu$J resulted in VUV pulses of $\sim$4~$\mu$J at the upper limit (efficiencies up to $\sim$10$\%$).

\begin{figure}[hbt]
         \centering
         \includegraphics[width=0.45\textwidth]{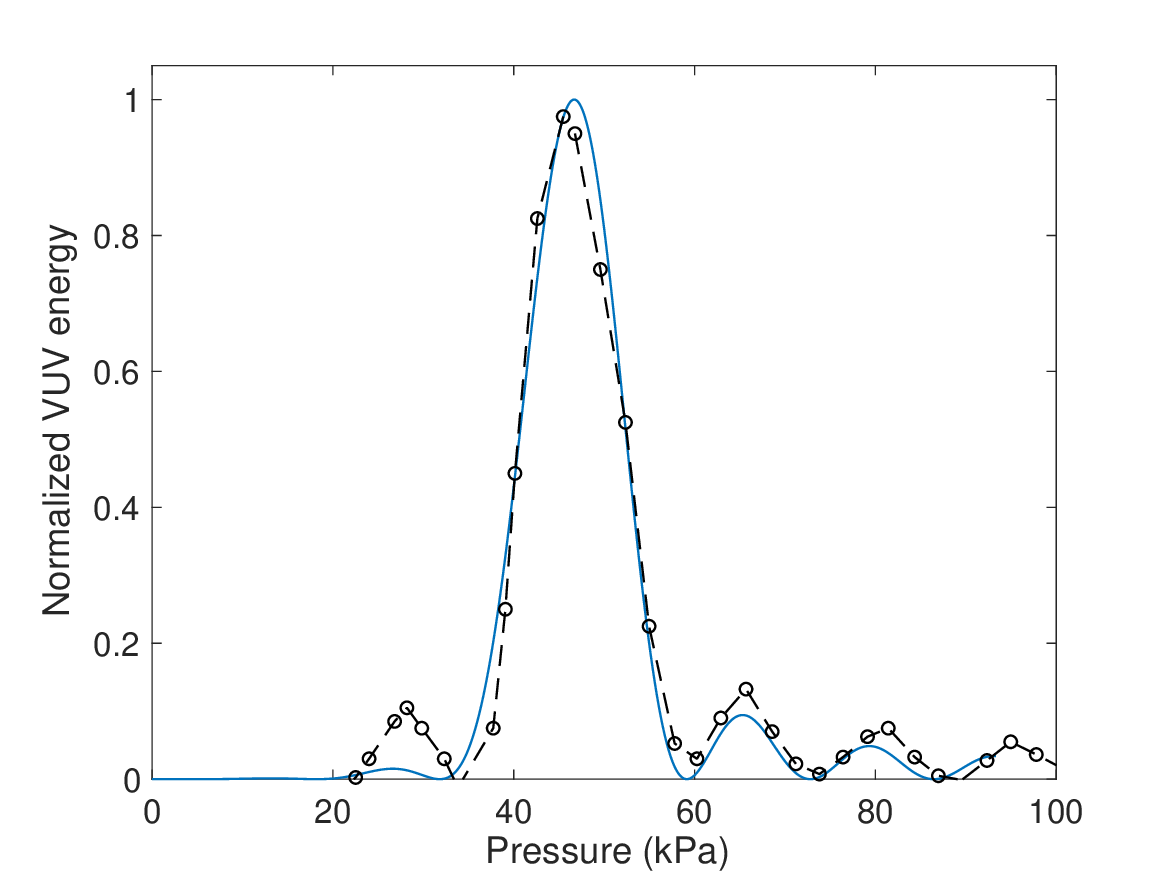}
         \caption{Experimental phase-matching curve (dashed line), with the simulation results from Eq.~\ref{eq:simple-PM} overlay (solid line). Both the calculated curve and the experimental data are normalized.}
         \label{fig:5w_phasematching_curve}
\end{figure}

As shown in Fig.~\ref{fig:5w_phasematching_curve}, the experimental pressure-dependence measurements also agree very well with the expectations from the simple theory. An optimized phase matching pressure of 46.7~kPa of He is found for the employed parameters. The theory phase-matching curve was based on Eq.~\ref{eq:simple-PM} above, using the Sellmeier formula for He from Ermolov {\it et al.} \cite{Ermolov2015}. This theory expression reproduces both the position and width of the main peak of the phase matching curve quite well, along with the smaller peaks at higher and lower pressures, expected from the $sinc^2(\Delta k L/2)$ phase-matching function. The experimental curve does show a slightly larger peak at $\sim$28.1~kPa. This feature may be caused by slight imperfection in the coupling of the pump, leading to phase matched generation of VUV in a higher order spatial mode at these lower pressures.

In order to be able to model the process in even more detail, and also include effects of cross-phase modulation (XPM), self-phase modulation (SPM), ionization, modal coupling and pump-depletion, we used Luna.jl~\cite{Luna}, a flexible platform for the simulation of nonlinear optical dynamics in waveguides, which has been shown to be quantitatively predictive against experiment \cite{Travers2019}. Using Luna.jl to generate a phase matching curve for the experimental conditions produces a curve that is virtually identical with the simple theory expression.
This observation is consistent with the expectation that at these pump/seed powers, the process is well described by the simple model, since there is no complicated SPM/XPM or plasma interaction at the entrance of the fiber, to initiate mode coupling, or modify the driving fields.

\begin{figure}[tbh]
         \centering
         \includegraphics[width=0.45\textwidth]{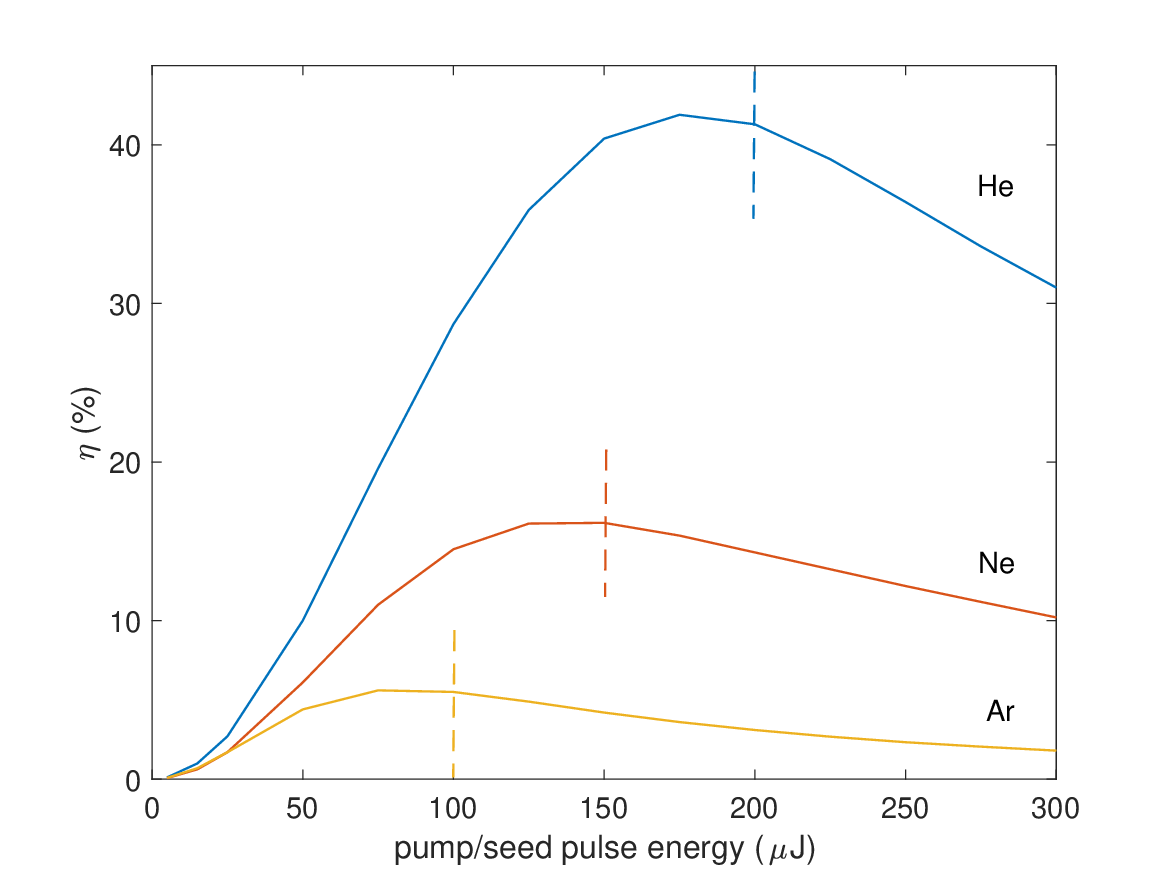}
         \caption{Conversion efficiency, $\eta$, of the 267~nm pump into VUV as a function of the pump/seed power for the fixed HCF geometry used, comparing He, Ne and Ar as the nonlinear medium. The dashed lines indicate the onset of mode-coupling, initiated by plasma formation at the entrance of the HCF.}
         \label{fig:Luna_5w_Optimal}
\end{figure} 

Exploring the power scaling beyond the current experimental limit, the Luna.jl simulations show no significant detrimental effects even when pumping/seeding with 125~$\mu$J (pump and seed energies are varied together in these simulations). At these powers, the simulations predict the generation of $\sim$45~$\mu$J of 160~nm, corresponding to a conversion efficiency of $\sim$36$\%$ when the process is optimally driven, see Fig.~\ref{fig:Luna_5w_Optimal}. From this point on, the conversion efficiency starts to roll over mainly due to plasma interaction, but a combination of XPM and pump-depletion can also be observed. As power is scaled higher still, the plasma formed at the entrance of the HCF, facilitating coupling of the EH$_{11}$  mode into higher order spatial modes (EH$_{12}$, EH$_{13}$ etc.), which propagate in the HCF with different group velocities, and can generate VUV in higher order spatial modes to the extent that the pressure allows these processes to phase-match. The onset of mode-coupling, defined as the point where 1$\%$ of the seed-pulse energy couples to higher order modes, is indicated by the dashed line at $\sim$200~$\mu$J.
 It is interesting to compare the different noble gasses as the nonlinear medium. Performing the same simulation for neon and argon, at their optimal phase matching pressure (22.0~kPa and 1.75~kPa, respectively) produces the results shown by the two other curves in Fig.~\ref{fig:Luna_5w_Optimal}.
The main observations when comparing He with Ne and Ar is that the intensities where the FWDFM process is optimally driven moves down, as does the maximal conversion efficiency ($\sim$40$\%$ for He, $\sim$15$\%$ for Ne and $\sim$5$\%$ for Ar). 
These simulations provide interesting insights into the more complicated optimally and strongly-driven regimes.
Presentation of full simulation results is beyond the scope of this letter, but results are available in the online data repository which accompanies this work \cite{fibreVUV2023dataRepo}.
\begin{figure}[t]
  \centering\begin{subfigure}[]{0.4\textwidth}
     \centering
     \includegraphics[width=\textwidth]{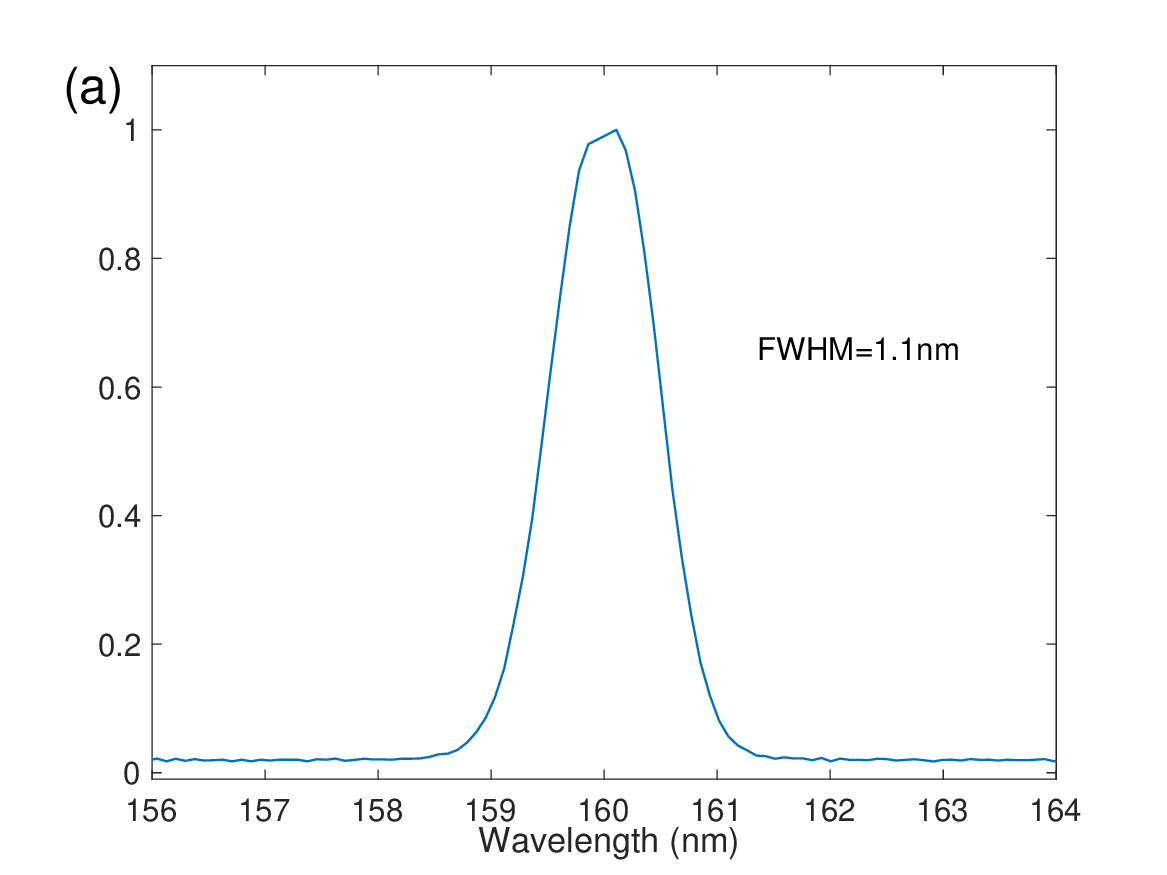}
    \end{subfigure}
    
    \begin{subfigure}[]{0.4\textwidth}
     \centering
     \includegraphics[width=\textwidth]{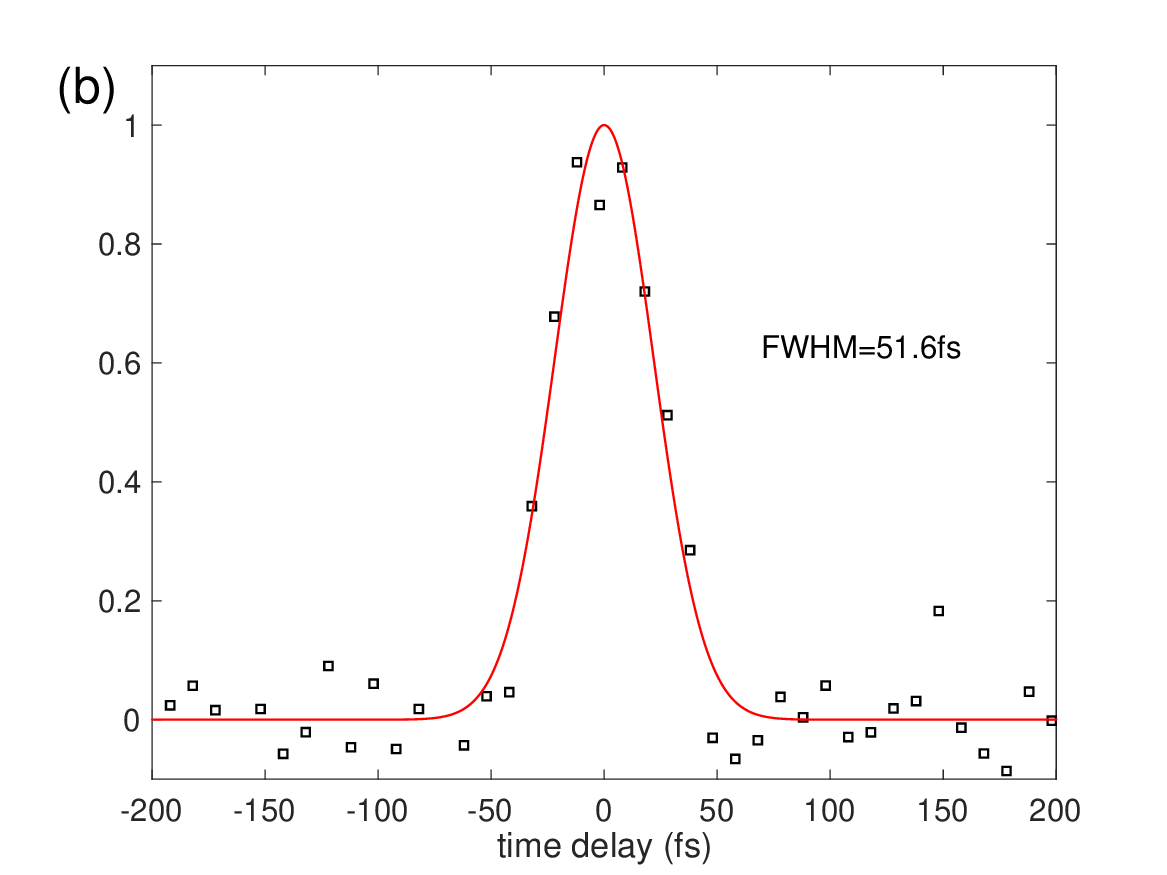}
   \end{subfigure}
    \caption{\label{fig:5w_spectrum_CC}(a) Spectrum of the generated VUV pulses at 160~nm, with an full width at half maximum (FWHM) of 1.1~nm, corresponding to a transform-limited pulse length of 35~fs. (b) Typical cross-correlation of the VUV pulse, performed in the ionization cell, see main text for details.}
\end{figure} 
The temporal and spectral characterization of the VUV pulses was performed by separate but concurrent measurements. A typical spectrum of the VUV pulse at 45.5~kPa of He is shown in the top panel of Fig.~\ref{fig:5w_spectrum_CC}. The full width at half maximum (FWHM) of the spectrum is $\sim$ 1.1~nm, which corresponds to a transform limited pulse length of 35~fs. The temporal measurement was performed by cross-correlation in a parallel plate ionization cell, filled with 1.1~kPa of Xe, where the VUV beam and a weak 3$\omega$ probe-beam, with a pulse length of 37~fs, were focused to the same spot, leading to non-resonant two-photon ionization when they overlap in time. The 3$\omega$ probe-pulse duration was calculated by determining the pulse duration after propagation of a 35~fs transform limited 3$\omega$ pulse through the 1~thick CaF$_2$ entrance window of the VUV chamber and the 0.5~mm thick MgF$_2$ input window of the ionization cell. At the powers used we observed no two-photon signal from the VUV pulse alone, but did have a background from 3-photon ionization by the 3$\omega$ probe-beam. Modulating the seed for the FWDFM process, and using lock-in detection at 500~Hz, cross-correlation traces could be obtained without signs of saturation. Typical FWHM of the cross-correlation traces were 52~fs, as can be seen in Fig.~\ref{fig:5w_spectrum_CC}. This corresponds to a VUV pulse length of 37~fs, given that the FWHM of the cross-correlation trace based on two-photon ionization can be expressed as $\tau_{\mathrm{cc}}^2=\tau_{3\omega}^2+\tau_{5\omega}^2$. This result is in excellent agreement with expectations, since propagation of a transform limited 35~fs pulse at 160~nm through the 0.5~mm MgF$_2$ input window of the ionization cell would stretch it to $\sim$~37~fs.

In conclusion, we have demonstrated the generation of near transform limited 37~fs VUV pulses at 160~nm, with pulse energies as high as 4~$\mu$J, which corresponds to a conversion efficiency of $\sim$10$\%$, rivaling the most efficient solid-state conversion schemes in non-linear crystals. The source is pumped by a commercial 1~kHz Ti:sapphire amplifier system, and uses a relatively simple experimental setup, taking advantage of phase-matched FWDFW in a He-filled stretched HCF. The separation of the VUV beam from the driving beams is done using dielectric mirrors. The high conversion efficiency of this source, makes it an ideal path to femtosecond tuneable VUV pulse generation, where standard OPAs with solid state frequency mixing schemes based on BBO could provide tunability from 185 to 140~nm (6.70-8.86~eV)~\cite{Noack_tuneableVUV_45fs,Trabs_2014,Forbes2021}.  
The combination of the brightness, short pulse duration, stability and high photon energy makes this source attractive for many forms of photon-in, electron-out spectroscopies~\cite{Trabs_2014,Forbes2018,Forbes2021}.  
The VUV pulse could be pre-compensated for propagation through the experimental chamber input window, by adding a positive phase to the seed ~\cite{Noack_12fs,Suzuki_Spectra_Phase_Transfer_UV}, and it should be possible to make even shorter VUV pulses by seeding with a broader spectrum \cite{Noack_12fs}. Finally, this technique also lends itself to the direct generation of VUV pulses in desired polarization states, e.g. circular, by controlling the polarization of the driving fields~\cite{Lekosiotis_2020}.






\begin{backmatter}
\bmsection{Funding} R.F. gratefully acknowledges support from the Linac Coherent Light Source, SLAC National Accelerator Laboratory, which is supported by
the US Department of Energy, Office of Science, Office of
Basic Energy Sciences, under contract no. DE-AC02-76SF00515. R.L., P.H. and Q.L. gratefully acknowledge support from the NRC-CSTIP Quantum Sensors grant (QSP-075-1).

\bmsection{Disclosures} The authors declare no conflicts of interest.

\bmsection{Data availability} Data underlying the results presented in this paper, along with additional Luna.jl simulations, are \href{https://dx.doi.org/10.6084/m9.figshare.24878970}{available in a Figshare data repository, DOI: 10.6084/m9.figshare.24878970} \cite{fibreVUV2023dataRepo}.
\end{backmatter}

\bigskip
\bibliography{sample, roadmaps_and_reviews_071223}
\bibliographyfullrefs{sample, roadmaps_and_reviews_071223}
%
%

\end{document}